\documentstyle[epsfig,rotate,12pt,here]{article}
\begin{document}
\centerline{\large {\bf STRUCTURAL RELAXATIONS}}
\centerline{\large {\bf  IN A SIMPLE MODEL MOLTEN SALT}}
\bigskip
\leftline{Matthias Fuchs}
\leftline{Department of Materials Science, University of Illinois, Urbana, Il 61801}

\bigskip\bigskip

\leftline{\large Abstract}
\bigskip

The structural relaxations of a dense, binary mixture of charged hard spheres are studied  
using the Mode Coupling Theory (MCT). Qualitative differences to non--ionic systems
are shown to result from the long--range Coulomb interaction and  charge
ordering in  dense molten salts. The presented non--equilibrium
results are determined by the equilibrium structure, which is input using the
well studied Mean Spherical Approximation.

\bigskip
\leftline{\large Introduction}
\bigskip

The equilibrium structure of ionic liquids is strongly affected by the 
long--range nature of the Coulomb interaction [1]. One aspect is the screening
of external charges familiar from the Debye-H\"uckel theory of ionic
solutions. Another aspect are oscillations in the charge density around
a given ion. These oscillations result from the competition between local
charge neutrality and the excluded volume restriction due to the finite 
diameter of the ions. Also the dynamics of equilibrium ionic liquids 
in the hydrodynamic regime differs from
the one of uncharged mixtures. Coulombic restoring forces lead  
to a non--diffusive and non--propagating relaxation of charge fluctuations [1].
The MCT  determines the slow structural relaxations of dense (supercooled) liquids 
from their equilibrium structure [2]. In this paper we discuss the most salient
features of these results arising from the long--range Coulombic interactions. 
These will be seen to be directly connected to the familiar effects
in the static structure mentioned above:  screening and charge ordering.

\bigskip
\leftline{\large Theory}
\bigskip

One of the most simple models of an molten salt is a binary mixture of hard spheres
with radii $d_i$ and charges $z_i$, $i=1,2$. Global charge neutrality fixes 
$z_1 \varrho_1 + z_2 \varrho_2 = 0$, where $\varrho_i$ denotes the density of species $i$. 
The mean spherical approximation [3] gives a satisfactory description of the equilibrium structure
of this system, which depends on the parameters $d_1/d_2$, $z_1/z_2$, packing fraction
$\varphi$ and coupling constant $\Gamma$. The packing fraction is the ratio of volume occupied by
spheres to the total volume. $\Gamma$ is a generalized inverse Debye screening length and a
measure of the Coulomb interaction compared to the thermal energy. At the values of these parameters
chosen in our study and collected in table I one observes a liquid to glass transition in the MCT
equations. This transition is the topic of the work reported.
It is worth noting that the density at the transition can be chosen to be lower than in 
the uncharged system [4]. Obviously, the charges increase the interactions of the particles. 
Asymmetric parameters ($d_1/d_2\ne1$) were first chosen in order to study experimentally 
more realistic non--symmetric salts. The charge asymmetry $z_1/z_2$, however, was then adjusted 
to obtain the value $\lambda=0.85$ for the exponent parameter $\lambda$; see the discussion.
The MCT for binary mixtures formulates a closed set of equations for the time and wave vector 
dependent density fluctuation functions, $F^{ab}_q(t) = \frac 1N \langle \delta\varrho^{a*}_q(t)
\delta\varrho^{b}_q(0) \rangle$ [5,2]. Of particular interest are the fluctuations of the
total-- or mass--, $\varrho^n=\varrho^1+\varrho^2$, and the charge--density,  
$\varrho^c=z_1\varrho^1+z_2\varrho^2$.
\begin{table}[H]
\centerline{Table I: Molten salt equilibrium parameters and results from MCT calculation.}
\begin{center}
\begin{tabular}[b]{*{5}{|c}|*{6}{|c}|}
\hline
$d_2/d_1$ & $z_1/z_2$ & $\varphi$ & $\Gamma$ & $k_BT\epsilon d_2/e^2$ & $\lambda$ & $b$ & 
$\gamma$ & $\sigma/\tau$ & $\varepsilon'_o$ & $\varepsilon'_\infty$ \\
\hline
1.2 & -3 & 0.475 & 1.68 & 0.113 \mbox{\footnotesize (from $\Gamma$)} 
& 0.845 & 0.40 & 3.24 & 0.49 & 14.9 & 3.56 \\
\hline
\end{tabular}
\end{center}
\end{table}
The initial values, $F_q^{ab}(t=0)=S^{ab}_q$, are the static structure factors which are the only
input determining the $F_q^{ab}(t)$ via the MCT equations. 

MCT shows that these equations exhibit
bifurcations which are identified as idealized liquid to glass transitions.  Close to these 
transitions the long time dynamics  is predicted to follow from
\begin{equation}
q^2 F_q(t) = S_q \{ M_q(t) S_q - \frac{d}{dt} \int_0^t dt' M_q(t-t') F_q(t') \}\; ,
\end{equation}
where matrix notation is used and the memory functions are quadratic polynomials in
the $F_q(t)$ correlators with coefficients determined by the static structure factors $S_q$ [2,5].
Thermally activated transport is neglected in (1) leading to its breakdown at low temperatures.
Only aspects which are not affected by this simplification will be discussed in this article.

\bigskip
\leftline{\large Results}
\bigskip

In order to screen an external charge the charge structure factor $S^{cc}_q$ has to vanish
for small $q$, $S^{cc}_q \propto (q\Lambda_D)^2$, where $\Lambda_D$ is the Debye 
screening length [1,3].
This leads to a decoupling of the MCT equations (1) for small wave vector. Whereas the
mass--density fluctuations are determined by a frequency dependent longitudinal viscosity 
$N^l(z=\omega+i\epsilon)$,
\begin{equation}
F^{nn}_q(z)/S^{nn}_q \to \frac{-1}{z-\frac{1}{S^{nn}_0 N^l_0(z)}}
\qquad \mbox{for $q\to 0$,  where }\qquad
N^{l}_q(z)= \frac{1}{q^2}  M^{nn}_q(z)\; ,
\end{equation}
the charge fluctuations couple to the generalized conductivity $\sigma(z)$ 
\begin{equation}
F^{cc}_q(z)/S^{cc}_q \to \frac{-1}{z+4\pi i \sigma(z)}
\qquad \mbox{for $q\to 0$,  where }\qquad
\sigma(z) = \lim_{q\to0} \, \frac{i}{4\pi} \frac{q^2}{S^{cc}_q M^{cc}_q(z)}\;  .
\end{equation}
These equations simplify further in two frequency windows reached close to
a transition. From the well known MCT results let us only mention the von Schweidler
decay which describes the onset of the $\alpha$--relaxation, i.e. the final decay into
equilibrium [2]:
\begin{equation}
F^{ab}_q(t)/S^{ab}_q  = f^{ab}_q - h^{ab}_q\; (t/\tau)^b \qquad\mbox{for intermediate times}
\; .
\end{equation}
The von Schweidler exponent $b$ and the exponent $\gamma$ determining the increase of the 
$\alpha$--relaxation time $\tau$ are functions of the exponent parameter $\lambda$
and uniquely determined at 
the chosen transition. Their values are included in table I. Eq. (4) shows that the density 
fluctuations exhibit a two--step relaxation. The amplitudes $f^{ab}_q$ of the final or 
$\alpha$--relaxation is smaller than unity. The relaxation is non--exponential in general and the
$\alpha$--relaxation times $\tau^{ab}_q$ depend sensitively on temperature or density 
and on wave vector by a multiplicative factor, which roughly equals $(f^{ab}_q/h^{ab}_q)^{1/b}$.
The mass--density correlation functions qualitatively agree with the results obtained
for neutral one--component liquids [2,6]. Fig. 1 shows the $\alpha$--amplitudes,
$f^{nn}_q$, which describe the frozen--in mass--density structure at the 
glass transition. 
\begin{figure}[H]
\centerline{\rotate[r]{\epsfysize=4.3in 
\epsffile{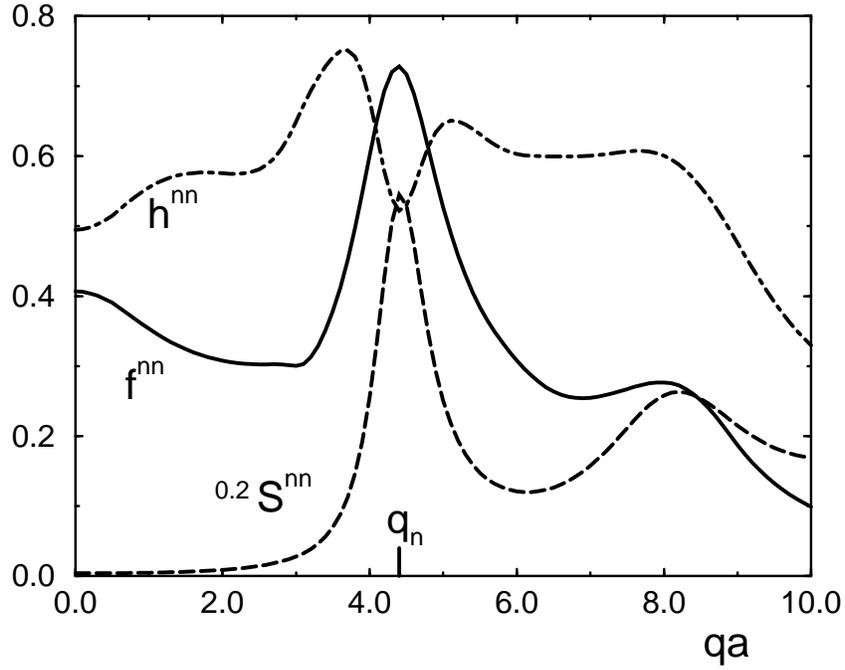}}}
\caption{Mass--density $\alpha$--amplitude $f^{nn}_q$, critical amplitude $h^{nn}_q$ and 
structure factor $S_q^{nn}$.
} 
\end{figure}
The known local packing on shells separated by the mean average 
interparticle spacing, $a\approx q_n/4.4$,
is seen in $f^{nn}_q$ as a  consequence of the one in $S^{nn}_q$ [2,6].
Local neutrality and excluded volume effects lead to charge ordering and a prominent peak 
in the charge structure 
factor, $S^{cc}_q$. The average spacing between ion--shells of equal sign is larger than the 
average particle spacing resulting in the peak in $S^{cc}_q$ to lie at $q_c$, where $q_c<q_n$. The
charge--density fluctuations $f^{cc}_q$, which are arrested at the transition, reflect this 
ordering [4]; see Fig. 2. 
\begin{figure}[H]
\centerline{\rotate[r]{\epsfysize=4.3in 
\epsffile{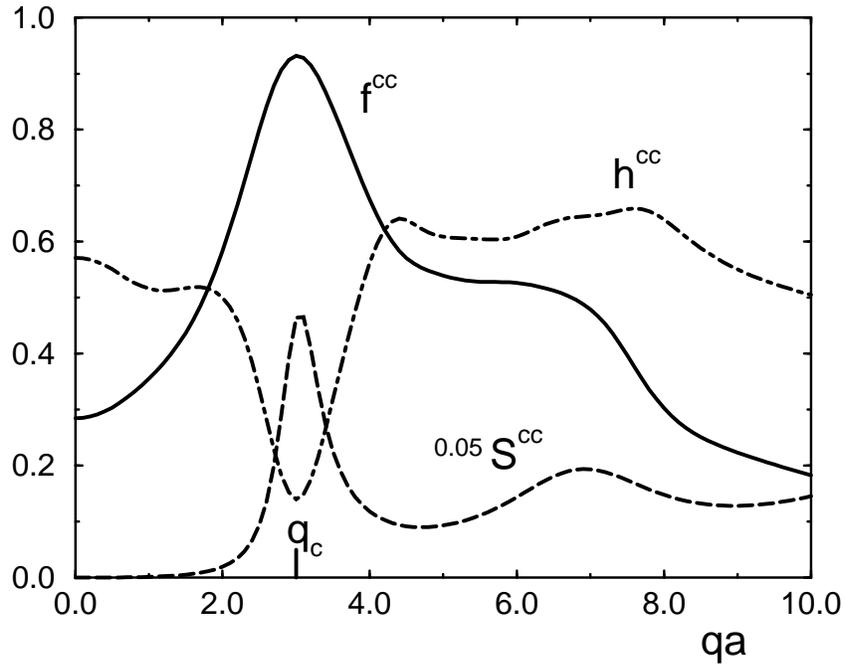}}}
\caption{Charge--density $\alpha$--amplitude $f^{cc}_q$, critical amplitude $h^{cc}_q$ and 
structure factor $S_q^{cc}$. 
} 
\end{figure}
The different mass-- and charge--density ordering also leads to specific variations in the
wave vector dependent prefactors of the $\alpha$--relaxation times $\tau^{nn}_q$ and $\tau^{cc}_q$.
Fig. 3 shows these variations as estimated from the peak positions in the corresponding
susceptibilities, $\tau\omega_{max}=1$. A $q$--dependent slowing down at the maximal amplitudes of the
$\alpha$--process is observed. Superficially, this mimics the known De Gennes narrowing, as the
$\tau_q$ vary in phase with the static structure factors [1]. 
However, this correlation only holds, because the $\alpha$--relaxation amplitudes, 
$f_q$, vary in phase and the critical amplitudes, $h_q$, vary out of phase with the structure factors,
$S_q$ [2]; see Figs. 1 and 2. 

The results presented thus far specify the structural relaxations on length scales of the order of
interparticle distances. Analyzing Eq. (1) more closely, it is seen that it  also is the 
static structure on these length scales which determines the results. The quantitative results 
therefore depend on the appropriateness
of the underlying microscopic model, i.e. the liquid of charged hard spheres with 
the choice of parameters. 
From Eqs. (1,3) one can also obtain macroscopic transport coefficients like the
conductivity $\sigma$ and the dielectric constant $\varepsilon'$ of the ionic melt. Their values are
included in table I. The frequency dependent conductivity determines the dielectric ``constant'' via
\begin{equation}
\varepsilon(z)=1+4\pi i\frac{\sigma(z)}{z}\; .
\end{equation}
The liquid molten salt is characterized by a conductivity and dielectric constant measured
at low frequencies, $\omega\tau \ll1$. In the idealized glassy state 
the particles are arrested and ionic transport over macroscopic distances is not possible.
The $\alpha$--relaxation time $\tau$ diverges and
the conductivity vanishes. Eq. (5) then results in a dielectric constant 
$\varepsilon_\infty$ which can also  be observed in the liquid state at high frequencies,
$\omega\tau\gg1$. Fig. 4 shows the dispersion of the dielectric constant versus
$\omega\tau$. In the same plot, the conductivity crosses over from its low
frequency value to a power law behavior, $\sigma(\omega\tau\gg1)\propto (\omega\tau)^{1-b}$,
at intermediate frequencies.    
\begin{figure}[H]
\centerline{\rotate[r]{\epsfysize=4.3in 
\epsffile{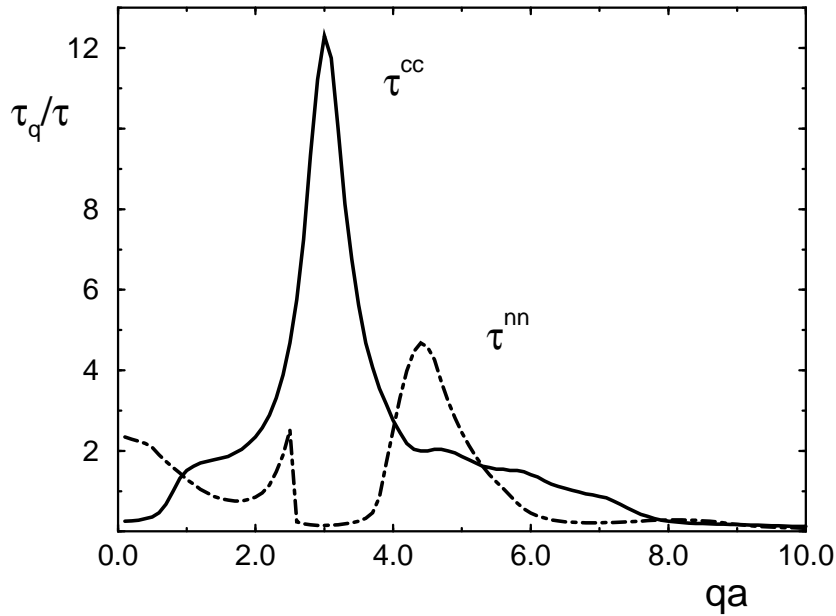}}}
\caption{Wave vector dependent factors of the $\alpha$--relaxation times $\tau^{nn}_q/\tau$ and
$\tau^{cc}_q/\tau$.
} 
\end{figure}
\newpage
\begin{figure}[H]
\centerline{\rotate[r]{\epsfysize=4in 
\epsffile{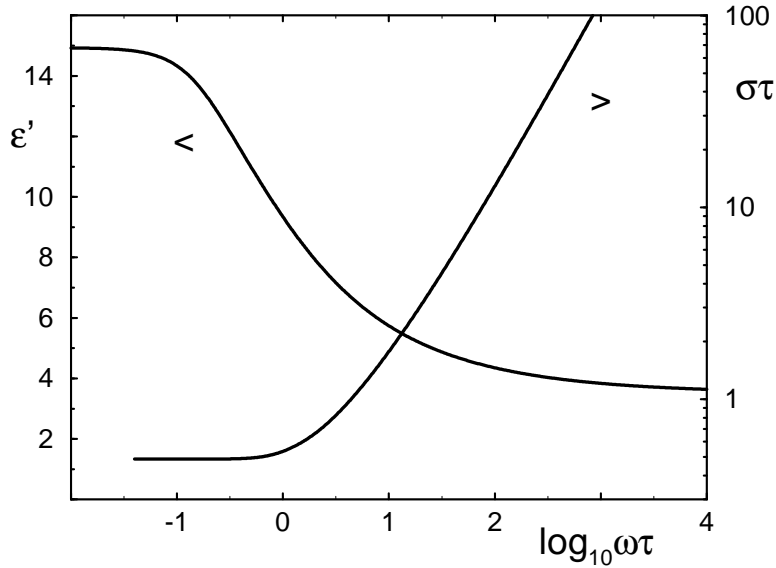}}}
\caption{Scaled conductivity $\sigma\tau$ and real part of the dielectric constant $\varepsilon$ 
versus $\omega\tau$. 
} 
\end{figure}

\bigskip
\leftline{\large Discussion}
\bigskip

The  mixed salt CKN [40\% Ca(NO$_3$)$_2$- 60\% KNO$_3$] is a well studied glassforming melt; see
[7,8] and references therein. Its dynamics in a wide temperature range has been studied by
neutron [7] and by  depolarized light scattering [8]. In the latter work it was found that the
exponent parameter $\lambda\approx$ 0.85 describes the dynamics in an intermediate time
window. These findings lead to the choice of parameters in our model reproducing this $\lambda$--value.
This assures that the asymptotic dynamics in the intermediate time window is described
by the correct asymptotic master function exhibiting, for example, the von Schweidler
asymptote of Eq. (4) or Fig. 4. 

The wave vector dependent prefactors, like $f_q$, $h_q$ and $\tau_q$,
are strongly coupled to the structural input as specified by our model and discussed in
the previous chapter. 
It cannot be expected that the simple model reproduces these amplitudes quantitatively. 
However, the variations of the $\alpha$--amplitude $f^{cc}_q$ and relaxation time $\tau^{cc}_q$ 
in phase with $S^{cc}_q$ and of the critical amplitude $h^{cc}_q$ out of phase with it, 
are expected to be general findings applicable to molten salts. The corresponding variations in
the mass--density quantities shown in Fig. 1 have  been found in 
MCT calculations for different simple liquids [2,6]. They have been compared to dynamic light
scattering spectra in colloidal suspensions; general agreement with errors of the order of 15\%
was observed [9]. Neutron scattering from CKN
measures a combination of charge-- and mass--density fluctuations determined by the different
atomic neutron scattering cross sections. Our findings of peaks in the $\alpha$--amplitudes
and in the $\alpha$--times at the wave vectors, $q_c$ and $q_n$, 
characterizing charge and density fluctuations  in the equilibrium 
structure, are in qualitative agreement with the reported measurements [7]. 
The results show that charge ordering and local packing are the underlying mechanism.

Let us restate, that the
increase of the time scales at peaks in the static structure factors is not a simple example of 
de Gennes narrowing and cannot be explained by short time sum rules but is a consequence of
the MCT equations  (1) [2]. Eqs. (1)  do not include
the short time dynamics and consequently violate the short time sum rules.
The shown 
variations result from the specific magnitudes of the coupling of different modes 
in the memory functions; the couplings are determined by $S_q$.

The MCT equations (1) correctly describe non--propagating and non--diffusive charge
fluctuations in the long--wavelength limit (3). This is a consequence of  screening in ionic melts
which requires finite restoring forces for charge fluctuations even on long wavelengths.
The generalized conductivity 
changes from a low frequency constant, $\sigma \propto 1/\tau$, to a power law behavior at large
$\omega\tau$.

In a cursory search liquid glass transitions were located for different parameters in this model.
The parameter of table I
lead to a rather strong peak in the charge structure factor, $S_q^{cc}$. The origin of this
is the  large charge asymmetry $z_2/z_2=$-3 which entails a corresponding concentration 
ratio due to the requirement of global neutrality. The overestimated charge oscillations
lead to two special features in our results. First, the maxima in the $q$--dependent
amplitudes of the charge fluctuations are quite pronounced. This prevents any quantitative
comparisons with the neutron or light scattering data of [7,8]. This failure emphasizes that
quantitative comparisons between MCT calculations for simple liquids and experimental data
require appropriate microscopic models if non--universal features of the theory are tested; see
[6,9]. Second, comparing Fig. 1 to the corresponding results for one--component liquids, shows 
that the frozen--in mass-density structure of the formed glass is similar  but somewhat
distorted. Inspection of other parameter values in this model
reveals that another glassy structure
becomes stable if the charge asymmetry is increased some more. The small stability of the
studied glass to  another amorphous structure is the origin of the large exponent parameter.
This second glass differs in the frozen--in mass--density but not in the charge--density structure.
A striking consequence of the proximity of the two glassy states is the splitting
of the $\alpha$--relaxation in $F^{nn}_q(t)$ for some wave vectors 
into two processes. This effect is indicated in Fig. 3, where the inverse $\alpha$--peak position
frequency jumps from one process to the other at $qa\approx$2.5. This phenomenon 
has been discussed in schematic MCT models [10] and will be studied further 
in this microscopic model. 

\bigskip
\leftline{\large Acknowledgments}
\bigskip
Financial support from the Deutsche Forschungsgemeinschaft under contract
Fu 309/1-1 is acknowledged.

\bigskip
\leftline{\large References}
\bigskip

\parindent0pt 1. J.P. Hansen and I.R. McDonald, {\it Theory
of Simple Liquids}, 2nd edn. (Academic Press, London, 1986), p.364.\newline
\parindent0pt 2. W. G\"otze and L. Sj\"ogren, Rep. Prog.
Phys. {\bf 55}, 241 (1992); and references therein.\newline
\parindent0pt 3. L. Blum and J.S. H{\o}ye, J.Phys.Chem. {\bf 81}, 1311 (1977).\newline
\parindent0pt 4. J.S. Thakur and J. Bosse, J.Non-Cryst.Solids {\bf 117/118}, 898 (1990).\newline
\parindent0pt 5. J. Bosse and T. Munakata, Phys.Rev. {\bf A 25}, 2763 (1982);
L. Sj\"ogren and F. Yoshida, J.Chem.Phys. {\bf 77}, 3703 (1982).\newline
\parindent0pt 6. M. Fuchs, I. Hofacker and A. Latz, Phys.Rev. {\bf A 45}, 898
(1992); and references therein.\newline
\parindent0pt 7. F. Mezei, W. Knaak and B. Farago, Physica Scripta {\bf T 49}, 363 (1987). \newline
\parindent0pt 8. H.Z. Cummins, W.M. Du, M. Fuchs, W. G\"otze, S. Hildebrand, A. Latz,
G. Li and N.J. Tao, Phys.Rev. {\bf E 47}, 4223 (1993).\newline
\parindent0pt 9. W. van Megen and S.M. Underwood, Phys.Rev. {\bf E 49}, 4206 (1994).\newline
\parindent0pt 10.  M. Fuchs, W. G\"otze, I. Hofacker and A. Latz, J.Phys.:Condens.Matter {\bf 3},
5047 (1991).
\end{document}